\documentclass[twocolumn,showpacs]{revtex4}
\usepackage{graphicx}
\usepackage{bm}

\begin{document}

\def\be{\begin{equation}}
\def\ee{\end{equation}}


\newcommand{\p}{\phi}

\title{Quintessence,
super-quintessence and
observable quantities in Brans-Dicke and
non-minimally coupled theories}
\author{Diego F. Torres }
\address{Physics Department, Princeton University, NJ 08544, USA}

\begin{abstract}
The different definitions for the equation of state of a
non-minimally coupled scalar field that have been introduced in
the literature are analyzed. Particular emphasis is made upon
those features that could yield to an observable way of
distinguishing non-minimally coupled theories from General
Relativity, with the same or with alternate potentials. It is
found that some earlier claims on that super-quintessence, a stage
of super-accelerated expansion of the universe, is possible within
realistic non-minimally coupled theories are the result of an
arguable definition of the equation of state. In particular, it is
shown that these previous results do not import any observable
consequence, i.e. that the theories are observationally identical
to General Relativity models and that super-quintessence is not
more than a mathematical outcome. Finally, in the case of
non-minimally coupled theories with coupling $F=1+\xi \phi^2$ and
tracking potentials, it is shown that no super-quintessence is
possible.
\end{abstract}
\pacs{98.80.Cq, 98.70.Vc} 
\maketitle

\section{Introduction}

During the last years, observations of distant Type Ia supernovae
\cite{1} and CMB measurements \cite{2} have shown that the
universe is, most likely, undergoing a process of accelerated
expansion. The widespread vision of the cosmological model is,
since then, a spatially flat low matter density universe. This
implies that the total energy density today is dominated by a
contribution having negative pressure (cosmological constant, or
quintessence \cite{Q}) which has just began to undertake the
leading role in the right hand side of the Einstein field
equations.

The cosmological constant {\it solution} to this state of affairs
appears not to be completely satisfactory (see for instance
\cite{3}). Precise initial conditions should be given in order to
solve the coincidence problem (why the vacuum energy is dominating
the energy density right now). Moreover, a fine-tuning problem
appears, since a vacuum energy density of order $\sim10^{-47}{\rm
GeV}^4$ requires a new mass scale about 14 orders of magnitude
smaller than the electroweak scale, having no a-priori reason to
exist. In addition, the equation of state, $p/\rho$, for vacuum
energy is exactly equal to $-1$, what at first sight appears as
yet another value which is, in a dynamical setting, precisely set.
Quintessence \cite{Q}, and its derived models, being the main
alternatives, are based on the existence of one or many scalar
fields, which dynamically evolve together with all others
components of the universe. The above-mentioned problems are
alleviated within this framework. A sub-class of models, those
having inverse power law potentials, present tracking solutions
where a given amount of scalar field energy density can be reached
starting from a large range of initial conditions (see for e.g.
Refs. \cite{tracking}).

The simplest models of quintessence are based on minimally coupled
scalar fields. For a general potential $V$, the equation of state
for those quintessence models is given by \be \frac {p}{\rho}=
\frac{\dot \phi ^2 - 2V}{\dot \phi ^2 + 2V},\ee and it can be
easily proven that this expression is bounded to be within the
range -1 $\leq p/\rho \leq$ 1, unless of course one is willing to
accept negative defined potentials. In the latter cases, the
energy density becomes itself a negative quantity. For usual
models of quintessence, then, it is clear that no
super-acceleration can appear. The latter is a result of an
extremely negative ($<-1$) equation of state. This possible
super-accelerated expansion has been recently dubbed
super-quintessence by several authors, e.g.  \cite{faraoni},
although its consequences are being analyzed since some time
before \cite{menace}. The main reason supporting this interest is
that current observational constraints are indeed compatible with,
if not favoring, such values for the equation of state (see for
instance Ref. \cite{menace}).

Extended quintessence models are those in which the underlying
theory of gravity contains a non-minimally coupled scalar field.
It is this same scalar field which, apart from participating in
the gravity sector of the theory, is enhanced by a potential to
fulfill the role of normal quintessence. From a theoretically
point of view, these ideas are appealing: it is the theory of
gravity itself what provides the dynamically evolving, and
currently dominating, field. Recent works on this area include
those presented in Refs.
\cite{faraoni,Sen,perrota,per2,uzan,farese,chiba}. We will have
the opportunity to comment with much more detail on some of these
works below.  In addition, just to quote a few others in a so
vastly covered topic, see the works of Ref. \cite{new}. We would
also like to remark that one of the first detailed analysis of a
non-minimally coupled theory with a scalar field potential was
made by Santos and Gregory \cite{santos}, years before the concept
of quintessence was introduced.

It has been claimed by many that a non-minimally coupled theory,
like for instance Brans-Dicke gravity, can harbor
super-quintessence solutions (e.g. \cite{faraoni,Sen}).  However,
there are different, and in most cases conflicting, definitions
for the equation of state in these theories. Then, care should be
exercised when analyzing the claims of the existence of
super-quintessence solutions: in some cases, they do not report
either any physical import, because the equation of state really
is not more than a complex relationship between the field and its
derivative without any supporting conservation law, nor any
observational consequence, because the amount of
super-quintessence is so small that is far beyond any foreseen
experiment. It is the aim of this paper to help in clarifying
these points, and to analyze, from an observational point of view,
how non-minimally coupled theories differentiate from usual
General Relativity in what concerns to quintessence and
super-quintessence models.

The rest of this work is presented as follows. In the following
Section we comment on the energy conditions and the status of
super-quintessence regarding them. Then, we introduce the gravity
theories we are interested in. Section IV analyzes the case of
Brans-Dicke gravity whereas Section V studies more general
non-minimally coupled theories. A discussion and summary of the
results is given in Section VI. A brief Appendix discusses an
alternative formulation of the theories of gravity, useful for
numerical computations.

\section{The energy conditions}

For a Friedman-Robertson-Walker space-time and a diagonal
stress-energy tensor $T_{\mu\nu}=(\rho,-p,-p,-p)$ with $\rho$
being the energy density and $p$ the pressure of the fluid, the
energy conditions (EC) read:
\begin{eqnarray}
\hbox{null: NEC} & \iff &  (\rho + p \geq 0 ), \nonumber \\
\hbox{weak: WEC} & \iff &  (\rho \geq 0 ) \hbox{ and } (\rho + p
\geq 0),  \nonumber \\ \hbox{strong: SEC} & \iff & (\rho +
3 p \geq 0 ) \hbox{ and } (\rho + p \geq 0), \nonumber \\
\hbox{dominant: DEC} & \iff &  (\rho \geq 0 ) \hbox{ and } (\rho
\pm p \geq 0).
\end{eqnarray}
They are, then, linear relationships between the energy density
and the pressure of the matter/fields generating the space-time
curvature. Violations of the EC have sometimes been presented as
only being produced by unphysical stress-energy tensors. If NEC is
violated, and then WEC is violated as well, negative energy
densities --and so negative masses-- are thus physically admitted.
However, although the EC are widely used to prove theorems
concerning singularities and black hole thermodynamics, such as
the area increase theorem, the topological censorship theorem, and
the singularity theorem of stellar collapse \cite{VISSER-BOOK},
they lack a rigorous proof from fundamental principles. Moreover,
several situations in which they are violated are known; perhaps
the most quoted being the Casimir effect, see for instance Refs.
\cite{VISSER-BOOK,last-visser} for additional discussion. Observed
violations are produced by small quantum systems, resulting of the
order of $\hbar$. It is currently far from clear whether there
could be macroscopic quantities of such an exotic, e.g.
WEC-violating, matter/fields may exist in the universe. A program
for imposing observational bounds (basically using gravitational
micro and macrolensing) on the existence of matter violating some
of the EC conditions has been already initiated, and experiments
are beginning to actively search for the predicted signatures
\cite{obs}. Wormhole solutions to the Einstein field equations,
extensively studied in the last decade (see Refs.
\cite{VISSER-BOOK,wh} for particular examples), violate the energy
conditions, particularly NEC. Wormholes are probably the most
interesting physical entity that could exist out of a macroscopic
violation of the EC.

It is interesting to analyze what does super-quintessence imply
concerning the validity of the EC. As stated in the Introduction,
super-quintessence is described by a cosmic equation of state \be
\frac p\rho < -1, \ee and so different situations arise depending
on the sign of the energy density $\rho$. If $\rho >0$,
super-quintessence implies $p + \rho < 0$, and thus the violation
of all the point-wise EC quoted above. Note that WEC is violated
because of the violation of its second inequality. If, on the
contrary, already $\rho < 0$, then NEC may be sustained, but WEC
is violated. Super-quintessence then implies strong violations of
the commonly cherished EC. But, should this be taken as
sufficiently unphysical as to discard a priori the possibility of
a super-accelerating phase of the universe?

Apart from the finally relevant response, coming from experiments
(today super-quintessence equations of state are not discarded,
and maybe even favored by experimental data, see for e.g.
\cite{menace}), the answer will of course rely on how much do we
trust the EC, which, as we have already said, are no more than
conjectures. Particularly for non-minimally coupled theories,
violations of the EC are much more common than in General
Relativity, see for instance the works of Ref. \cite{last-visser}
and references therein. In addition, recently, the consequences of
the energy conditions were confronted with possible values of the
Hubble parameter and the gravitational redshifts of the oldest
stars in the galactic halo \cite{SEC}. It was deduced that for the
currently favored values of $H_0$, the strong energy condition
should have been violated sometime between the formation of the
oldest stars and the present epoch. SEC violation may or may not
imply the violation of the more basic EC, i.e. NEC and WEC,
something that have been impossible yet to determine. In any case,
super-quintessence could be a nice theoretical framework for
explaining observational data opposing the EC. To the study of
super-quintessence in non-minimally coupled theories, we devote
the rest of this paper.

\section{Gravity theory}

In this section we shall present the general non-minimally coupled
Lagrangian density given by
\begin{eqnarray}
\label{action} S=\int d^4 x \sqrt{-g} \left[ {1 \over 2} f(\phi,
R) - {\omega (\phi )\over 2} \nabla^{\mu} \phi \nabla_{\mu}\phi
\right. \nonumber \\ \left. -V( \phi) + L_{fluid}\right]\ .
\end{eqnarray}
Here, $R$ is the Ricci scalar and units are chosen such that
$8\pi G=1$. The functions $\omega (\phi)$ and $V(\phi )$ specify the
kinetic and potential scalar field energies, respectively. The
Lagrangian $L_{fluid}$ includes all the components but $\phi$.
The function $f$ will be assumed to be of the form
\be
f(\phi,R)=F(\phi)\,R.\ee Einstein equations from the
general action (\ref{action}) are:
\begin{equation}
{H}^2= {1 \over 3F} \left( {\rho}_{fluid} +{\omega \over 2}
\dot{\phi}^2 +V - 3 { H} \dot{F} \right) \;,
\end{equation}
\begin{equation}
\dot{ H}=- {1 \over 2F} \left[ ({\rho}_{fluid}+p_{fluid}) +\omega
\dot{\phi}^2 + \ddot{F} - { H} \dot{F}  \right],
\end{equation}
\begin{equation}
\ddot{\phi}+3{H} \dot{\phi}=-{1 \over 2 \omega} \left(
 \omega_{, \phi} \dot{\phi}^2 - F_{, \phi} R +2 V_{, \phi} \right)
 \;,
\end{equation}
where overdots denote normal time derivatives. The Klein-Gordon
equation is actually very complicated in the general case. Using that
\begin{eqnarray} && R= 6 (\dot H + 2 H^2)=\frac 1F
\nonumber \\ && \left[ \rho_{fluid} - 3p_{fluid}
-\omega \dot \phi^2 + 4 V - 3 (\ddot F + 3 H \dot F) \right],  \end{eqnarray}
after
some algebra, it ends up being
\begin{eqnarray} && \!\!\!\!\!\!\! \ddot \phi \left(1+ \frac{3
F_{,\,\phi}^2}{2\omega F} \right) = -3 H \dot \phi - \frac
1{2\omega} \omega_{,\,\phi} \dot \phi^2 -\frac 1\omega
V_{,\,\phi}-    \frac{F_{,\,\phi}}{2F\omega} \nonumber \\ &&
(-\rho_{fluid}+3p_{fluid} )- \frac{F_{,\,\phi}}{2F} \dot \phi^2
+\frac{F_{,\,\phi}}{\omega F} 2V - \nonumber \\
&&\frac{3F_{,\,\phi}}{2\omega F}F_{,\,\phi\,\phi}\dot \phi^2 -
\frac{F_{,\,\phi}^2}{2\omega F} 9 H \dot \phi . \end{eqnarray}

Two different kinds of theories are usually studied. One is the
archetypical Brans-Dicke gravity \cite{BD}, which appears by
choosing $\omega = \omega_{BD}/\phi$ and $F = \phi$, where
$\omega_{BD}$ is referred to as the coupling parameter. The other,
generically named as non-minimally coupled theories (although of
course Brans-Dicke gravity also has a non-minimally coupled scalar
field), are those for which $\omega=1$, and $F$ and the potential
$V$ are generic functions of the field. Interesting differences appear
when in the latter cases is of the form $F=const. + g(\phi)$, they will
be discussed below. At least formally,
starting from one of these Lagrangian densities, one can always
rephrase it into the alternative form by a redefinition of the
scalar field. Sometimes, however, this cannot be achieved with
closed analytical formulae.

\subsection{Experimental constraints}

The predictions of General Relativity in the weak field limit are
confirmed within less than 1\% \cite{limits2-cons}. Any
scalar-tensor gravity theory, then, should produce predictions
that deviate from those of GR by less than this amount in the
current cosmological era. In general, these deviations from GR can
be specified by the post-Newtonian parameters \cite{WILL} \be
\gamma -1 = - \frac{(dF/d\phi)^2}{\omega F + (dF/d\phi)^2},\ee \be
\beta - 1 = \frac 14 \frac{F(dF/d\phi)}{2\omega F + 3(dF/d\phi)^2}
\frac{d\gamma}{d\phi}.\ee Solar system tests currently constrain
\cite{limits2-cons}: \be |\gamma -1 | < 2 \;10^{-3},
\hspace{0.5cm} |\beta -1 | < 6 \; 10^{-4}, \ee  and they translate
into a limit on $1/F(dF/d\phi)^2$ at the current time, supposing
$\omega=1$, specifically  $1/F(dF/d\phi)^2 < 2\; 10^{-3}$
\cite{uzan}. If on the contrary, we assume the form of Brans-Dicke
theory, they imply $\omega_{BD}>500$. This value has been derived
from timing experiments using the Viking space probe \cite{limit}.
In other situations, claims have been made to increase this lower
limit up to several thousands, see Ref. \cite{limits2-cons} for a
review.

Starting from the action, one can define the cosmological
gravitational constant as $1/F$. This factor, however, does not
have the same meaning than the Newton gravitational constant of
GR. The Newtonian force measured in Cavendish-type experiments
between two masses $m_1$ and $m_2$ separated a distance $r$ is
$G_{{\rm eff}} m_1 m_2 / r^2$, where $G_{{\rm eff}}$ is given by
\cite{WILL}
\begin{equation}
\label{f8} G_{{\rm eff}}={1\over F} \left({2 \omega
F+4F_{,\phi}^{2}\over 2\omega F+3F_{,\phi}^{2}}\right)\ .
\end{equation}
The previous expression reduces to the well-known equality
$G_{{\rm eff}}(t)=(1/{\phi(t)})(2\omega+4)/({2\omega+3})$ for
Brans-Dicke theory. Current constraints imply \be |\frac {\dot
G_{{\rm eff}}}{G_{{\rm eff}}}| < 6 \; 10^{-12} {\rm yr}^{-1}  .
\ee In general, though, one cannot make the statement that this
constraint does directly translate into one for $\dot F/F$, for
one could in principle find a theory for which even when $F$
varies significantly, $G_{{\rm eff}}$ does not. Example of this is
the case of Barker's theory \cite{BARKER}, where $G_{{\rm eff}}$
is strictly constant.

Nucleosynthesis constraints can also be set for $\omega$, however
their impact is smaller than those set up in current experiments
(see for e.g. Refs. \cite{limits3-nucleo} and articles quoted
therein).

\subsection{The General Relativity limit}

There are important differences between Brans-Dicke gravity and
more general non-minimally coupled theories, particularly in what
refers to quintessence.

In the case of Brans-Dicke, when the coupling parameter is large,
the field decouples from gravity, and the theory reduces itself to
General Relativity \footnote{Strictly speaking, this seems to be
true only in the cases in which the trace of the energy-momentum
tensor for normal matter fields is not zero (see \cite{strange}
and references therein).}. When there is a potential, the limit of
$\omega_{BD} \rightarrow \infty$ would make the theory GR +
$\Lambda$ for every $V$. This is certainly not the case in
non-minimally coupled theories when $F$ involves a term
independent of the field (a constant). The limiting case of a
non-variable $F$-function is, in that situation, not GR plus a
cosmological constant, but GR plus the same potential. In this
case, then, the field recovers the status of normal quintessence,
being it minimally coupled and enhanced by a generic potential. It
is only in this sense that comparing different theories with the
same potential is justified. The same procedure do not provide
meaningful results when working with Brans-Dicke (or induced
gravity) models. To see how this difference appears it would be
enough to focus on the different Klein-Gordon equations for both
theories. In the case of Brans-Dicke, \be {\ddot{\phi}}+3{\dot
a\over{a}}{\dot\phi}={(\rho-3p)\over{2\omega_{BD}+3}}+
{2\over{2\omega_{BD}+3}}\left[2V-\phi{dV\over{d\phi}}\right], \ee
and all terms in the right hand side (including those having the
potential) are proportional to $1/\omega_{BD}$. Then, a
sufficiently large value of $\omega_{BD}$ will make this equation
sourceless, i.e. a solution being $\phi=const.$,
and reduce any $V$ in the Lagragian to a constant as well.
This does not happen in the case of non-minimally coupled theories
where $F$ contains an independent factor. For instance, in the
case in which $F=1+\xi \phi^2$, the Klein-Gordon equation is \be
\ddot \phi + 3 H \dot \phi  = \xi R \phi + \frac{dV}{d\phi},\ee
what clearly shows that the limit $\xi \rightarrow 0$ converts the
theory into normal quintessence.

\section{Brans-Dicke Theory}
\subsection{Field equations}

Consider the Brans-Dicke action given by
\begin{equation}
{\cal S}=\frac{1}{2} \int d^4x \sqrt{-g}\left[ \phi
R-{\omega(\phi)\over{\phi}}\phi^\alpha\phi_\alpha - 2
V(\phi)\right]+ L_{fluid} .
\end{equation}
 We shall consider that the
matter content of the universe is composed by one or several
(non-interacting) perfect fluids with stress energy tensor given
by
\begin{equation}
T_{\mu\nu}=(\rho+p){\it{v}}_\mu{\it{v}}_\nu+p
g_{\mu\nu},
\end{equation}
where ${\it{v}}_\mu{\it{v}}^\mu=-1$. This previous equation, then,
with adequate values of $\rho$ and $p$ will be valid for the
contributions of both, dust and radiation. Finally, we shall
assume that the universe is isotropic, homogeneous, and spatially
flat, and then represented by a $k=0$ Friedmann-Robertson-Walker
model whose metric reads
\begin{equation}
ds^2=-dt^2+a^2(t)[dr^2+r^2 d\theta^2+r^2\sin^2\theta d\phi^2].
\end{equation}
In this setting, the field equations are given by
\begin{equation}
\label{f1} {\dot a^2\over{a^2}}+{\dot
a\over{a}}{\dot\phi\over{\phi}}-{\omega\over{6}}
{\dot\phi^2\over{\phi^2}}-{ V\over{3\phi}}={\rho\over{3\phi}},
\end{equation}
\begin{equation}
\label{f2} 2{\ddot{a}\over{a}}+{\dot a^2\over{a^2}}+{\ddot
{\phi}\over{\phi}}+2{\dot
a\over{a}}{\dot\phi\over{\phi}}+{\omega\over{2}}
{\dot\phi^2\over{\phi^2}}-{ V\over{\phi}}=-{ p\over{\phi}},
\end{equation}
\be \label{f3}  {\ddot{\phi}}+3{\dot
a\over{a}}{\dot\phi}={(\rho-3p)\over{2\omega+3}}+
{2\over{2\omega+3}}\left[2V-\phi{dV\over{d\phi}}\right] . \ee To
simplify the notation in this Section we shall name $\omega_{BD}$
simply $\omega$. The continuity equation follows from the Bianchi
identity, yielding the usual relation
\begin{equation}
\label{f4} \dot\rho+3{\dot a\over{a}}(\rho+p)=0,
\end{equation}
which applied to both, matter ($p_m=0$) and radiation
($p_r=1/3\;\rho$), gives the standard dependencies:
\begin{equation}
\label{f5} \rho_m=\rho_{m,0} a^{-3}, \hspace{1cm}
\rho_r=\rho_{r,0} a^{-4}.
\end{equation}
We have chosen the scale factor normalization such that $a$ at the
present time is $a_0=1$. The current values of the densities are
given, in turn, by \be
\rho_{m,0}={3H_0^2}\Omega_{m,0},\hspace{1cm}
\rho_{r,0}={3H_0^2}\Omega_{r,0}.\ee Here, $H_0=100\times h$
km/s/Mpc is the current value of the Hubble parameter and
$\Omega_r=4.17\times 10^{-5}/h^2$ is the radiation contribution to
the critical density (taking into account both photons and
neutrinos, see \cite{Liddle-Lyth}). Typically, we shall work in a
model with $\Omega_m=0.4$, but this can be fixed to any other
value we wish, by using the contribution of the Brans-Dicke field
to respect the flatness of the universe.

The contribution of the field to the field equations can be
directly read from the field equations, if we replace the usual
General Relativity gravitational constant with the inverse of
$\phi$. The {\it effective energy and pressure} for the field end
up being,
\begin{equation}
\label{f6} \rho_\phi=3
\left[{\omega\over{6}}~{\dot\phi^2\over{\phi}}+{ V\over{3}}-{\dot
a\over{a}}\dot\phi\right],
\end{equation}
and
\begin{equation}
\label{f7} p_\phi=\left[{\omega\over{2}}~{\dot\phi^2\over{\phi}}-{
V}+\ddot\phi +2{\dot a\over{a}}\dot\phi\right].
\end{equation}

\subsection{Numerical implementation}

After being unable to find any obvious coordinates/field
transformation, in the sense explored by Mimoso \& Wands
\cite{mimoso}, Barrow \& Mimoso and Barrow \& Parsons
\cite{barrow}, and Torres \& Vucetich \cite{Torres:1996hv}, that
can deal with the complexities introduced by the appearance of the
self-interaction and solve the system analytically, we have
prepared a computed code to integrate the system
(\ref{f1}-\ref{f4}) numerically. Indeed, not all 4 equations in
the system are independent, because of the Bianchi identities, and
we have chosen to integrate Eqs. (\ref{f1}) and (\ref{f3}) having
as input the form of the matter densities given in Eq. (\ref{f5}).
We have followed the original idea of Brans-Dicke \cite{BD,Weim},
and transformed Eq. (\ref{f1}) into an equation for $H$, by
completing the binomial in the left hand side. Our variable of
integration was $x=\ln (t)$, and the output were $\ln (a)$, $\phi$
and $\phi^\prime$, where a prime denotes derivative with respect
to $x$. Having these values for each moment of the universe
evolution, it is immediate to obtain $\dot \phi$, $H$, and any
other quantity depending on them, like the effective pressure and
density of the Brans-Dicke field given in Eqs.
(\ref{f6}-\ref{f7}). The relevant initial condition of the
integration (in $\phi$) is chosen such that we fulfill today the
observational constraint ($G_{{\rm eff}}=1$) given by Eq.
(\ref{f8}). The derivative of the field can be set within a very
large range at the beginning since, while producing unobservable
changes at the early stages of the universe, this initial
condition is washed out by the evolution (a large range of
different initial conditions will give the same results). The
potential is generically written as \be
V=V=V_0\,f(\phi),\label{f9}\ee and the value of the constant $V_0$
is iteratively chosen such that it fulfills the requirement of a
large (say, $\Omega_\phi\sim0.6$) field contribution to the
critical density at the present time, for any given function
$f(\phi)$. We have tested our code in the limiting cases of the
problem and found agreement with previous results. As we have
discussed, when $\omega \rightarrow \infty$, Brans-Dicke theory
becomes General Relativity, $\phi$ being a constant, and every
potential effectively behave as a cosmological constant (i.e.
$p_\phi/\rho_\phi=-1$) during all the universe evolution.
Additionally, when we are in pure Brans-Dicke theory, without any
potential, we reproduce the results of Mazumdar et al.
 for the ratio between the Hubble length at equality and
the present one, $a_{eq}H_{eq}/H_0$ \cite{maz}.

\subsection{A worked example: $V=V_0\,\phi^{-2}$}

\begin{figure}
\includegraphics[width=9cm,height=12cm]{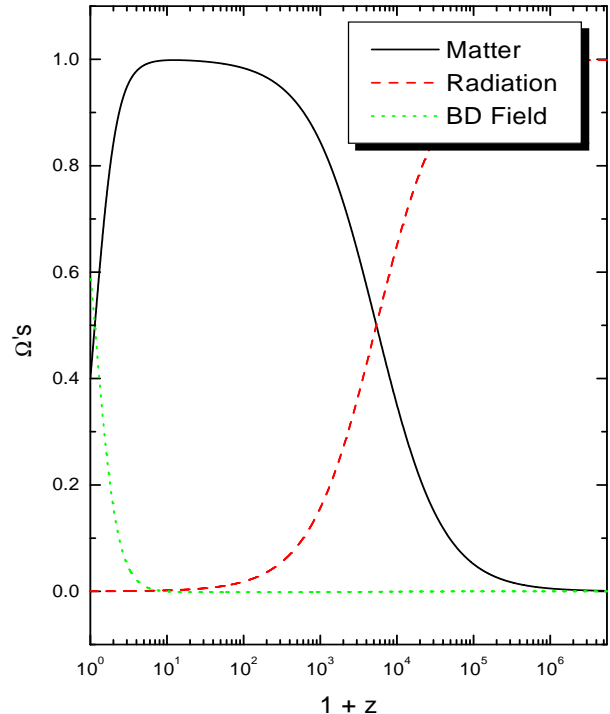}
\caption{Evolution of the contributions to the critical density
for an inverse square potential in a Brans-Dicke theory with
$\omega=500$. $H_0=75$ km/s/Mpc and $\Omega_{m,0}=0.4$. }
\label{fig1a}
\end{figure}

In Figure \ref{fig1a} we show the different contributions to the
critical density during the universe evolution for a Brans-Dicke
theory with $\omega=500$ and inverse square potential. The
contribution of the field is given, at any time of the universe
history, by \be \Omega_\phi=\frac{\rho_\phi}{3H^2\phi},\ee and the
others $\Omega$-values are defined in the same usual way as well.
We see that the Brans-Dicke field can act as quintessence, in
agreement with what other authors have previously found (see for
example Refs. \cite{perrota,chiba} and references therein). The
value of $V_0$ is extremely small, and mimics a cosmological
constant in General Relativity. The Brans-Dicke field and its
derivative do evolve in time. However, the current value of is
$\dot \phi = 9.8 \times 10^{-14}$ yr$^{-1}$, fulfilling the
above-mentioned constraint. The equivalence time (i.e. when
$\rho_{matter}=\rho_{radiation}$) in this model happens at 19801
yr, or $\ln (a)=-8.59$.

\begin{figure}
\includegraphics[width=9cm,height=12cm]{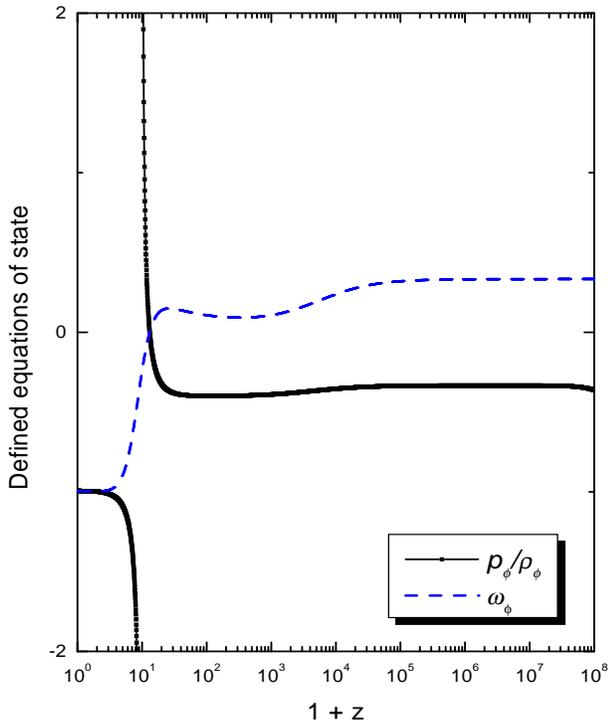}
\caption{Evolution of the ratio between the effective pressure and
densities for the Brans-Dicke field ($p_\phi/\rho_\phi$), and of
the effective equation of state for the field ($w_\phi$), entering
in the observable quantity $H$. The latter quantity is defined in
Section IID. Model parameters are as in Figure 1. } \label{bd-eos}
\end{figure}

In Figure \ref{bd-eos} we show, for the same model, the evolution
of the ratio between the effective pressure and densities for the
Brans-Dicke field. This ratio evolves strongly during the recent
matter era, the reason being that $\rho_\phi$ actually crosses
zero (from negative to positive values). This is in agreement with
Figure \ref{fig1a}, which shows the current field domination.\\

Indeed, we can obtain similar results to those presented here
changing the form of the potential to a variety of functional
dependencies on the Brans-Dicke field (see Table I). Most
interestingly, we see that at the present age of the universe, the
effective equation of state for some potentials  [we remark here
that this is an abuse of language, as will be explained below] is
smaller than -1.  As we stated, the phenomenon of having
$p_\phi/\rho_\phi<-1$ has been referred to as `super-quintessence'
by other authors, whereas the corresponding dark energy has been
dubbed `phantom energy' \cite{faraoni}. Apparently, already the
simplest potentials one can imagine can be super-quintessence
potentials within Brans-Dicke theory. It is true, however, that
the amount of super-quintessence we have found (how large is
deviation from -1 towards smaller values) is very small. It is
indeed much smaller than what other authors claimed before (see in
particular Ref. \cite{Sen}). However, we note that, apparently,
there is a sign mistake in their Klein-Gordon equation (7), the
last term in their rhs should be positive, what can be tested by
differentiation or comparison with, for example, Eq. (2.3) of Ref.
\cite{farese} or Eq. (2.6) of \cite{22} This can actually produce
a much bigger difference from an equation of state equal to -1 (as
we numerically tested), and is probably the origin of the
discrepancy.

As we have briefly implied above, $p_\phi/\rho_\phi<-1$ do not
have a especially clear physical meaning. Both, $p_\phi$ and
$\rho_\phi$ are made up of terms coming from the Lagrangian
density for the field. But they also contain terms coming from the
interaction between the field and gravity (through its non-minimal
coupling). The crucial aspect, then, is that the ratio
$p_\phi/\rho_\phi$ does not represent an equation of state, like
those of the other components, since an equation of the form $\dot
\rho_\phi + 3 \dot a /a (\rho_\phi+p_\phi)=0$ is not fulfilled. We
can actually see from first principles why $\dot \rho_\phi + 3
\dot a /a (\rho_\phi+p_\phi)\neq 0$ . The field equations of
Brans-Dicke theory in a general metric are \be \label{f11}
G_{\mu\nu}=\frac{1}{\phi}\left( T_{\mu\nu}^{matter} +
T_{\mu\nu}^{\phi}\right),\ee where $G_{\mu\nu}$ is the Einstein
tensor, $T_{\mu\nu}^{matter}$ is the stress-energy tensor for the
matter sector of the theory and \be  T_{\mu\nu}^{\phi}= \frac
w\phi \left( \phi_{,\mu}\phi_{,\nu}-\frac 12
g_{\mu\nu}\phi_{,\alpha}
\phi^{,\alpha}\right)+\left(\phi_{,\mu;\nu}-g_{\mu\nu} \Delta \phi
\right),\ee with $\Delta$ being the D$'$Alambertian operator. Now,
if we multiply the previous equation (\ref{f11}) by $\phi$ and
take covariant derivatives, it can be seen that \cite{BD} \be
T^{\mu\nu\;\;\phi}_{;\nu}= \left( \phi G_{\mu \nu}
\right)_{;\nu}\ee so that, whereas the usual continuity equation
for matter is valid, the continuity equation for the above-defined
`stress-energy tensor' for the field gets complicated.

\subsection{Observable quantities and the equation of state}

Then, if not $p_\phi/\rho_\phi$, which is the relevant (physically
meaningful) quantity to be considered as the equation of state for
the field in Brans-Dicke theory? {\it We suggest that the
important quantity to look at should come from what we actually
measure.} In the case of the homogeneous problem we are analyzing,
this is the Hubble parameter, $H$. If we rewrite the first of the
Friedmann-Robertson-Walker equations as \begin{eqnarray}
\label{f13} H^2&=&\frac 13 \left( \rho_{fluid} + \tilde
\rho_\phi\right)\nonumber \\
&=&\frac 1{3} \left(\rho_{m,0} a^{-3} + \rho_{r,0} a^{-4}+\tilde
\rho_{\phi\,0} a^{-3(1+w_\phi)}\right),\end{eqnarray} then,
$w_\phi$ is what is going to establish the departure from the
predictions of General Relativity plus cosmological constant or a
generic quintessence potential of a minimally coupled field. To be
specific, if $w_\phi=-1$, the theory would be indistinguishable
from General Relativity plus cosmological constant (from an
observational point of view). If $w_\phi>-1$, then the theory
would be indistinguishable from a normal (minimally coupled)
scalar field with a given potential. And finally, only if
$w_\phi<-1$, the theory would be observationally different from
its general relativistic counterparts: $w_\phi<-1$ is a value that
cannot be attained by any minimally coupled potential, as we
discussed in the Introduction. In that case, super-quintessence
adopts its rightful meaning: a super-accelerated expansion of the
universe, contrary to the case in which it just represent a
particular relationship between quantities importing no physically
clear concept. In the previous equation, and to be consistent with
the generalized field equations of the theory, we have defined
\begin{eqnarray} &&\;\;\tilde \rho= \tilde \rho_0 a^{-3(1+w_\phi)}=
-\rho_{m,0} a^{-3} -\nonumber \\ &&\rho_{r,0} a^{-4}+\frac1\phi
\left(\rho_{m,0} a^{-3} + \rho_{r,0}
a^{-4}+\rho_\phi\right),\end{eqnarray} where sub-indices $m,r$ and
0 stand for matter, radiation, and current values, respectively.
By definition, \be \label{extra} w_\phi=-1-\frac 13 \ln
\left(\frac{\tilde \rho_\phi}{\tilde\rho_{\phi\,0}}\right)\frac
1{\ln (a)}.\ee In Figure \ref{bd-eos} we have already shown the
evolution of this quantity in time (redshift). We also see that,
although some non-minimally coupled Brans-Dicke field can
effectively produce super-quintessence when supported by
particular potentials, the effective `equation of state' being
less than -1, its value is too close to -1 so as to effectively
mimics a cosmological constant at the current epoch. In Table I we
show that the same, almost imperceptible, deviation from an
effective cosmological constant appears for other potentials. This
is then showing a sign of caution when analyzing the impact of
non-minimally coupled theories using the equation of state:
differences are actually too small and fall below the threshold of
any current or foreseen experiment.

\begin{table}[t]\begin{center}
\caption{Values of the effective and phenomenological `equations
of state' for the Brans-Dicke field for different potentials.
Model parameters are $\omega=500$, $\Omega_r=4.17\times
10^{-5}/h^2$, $\Omega_m=0.4$, and $H_0=75$ km/s/Mpc.
 }\vspace{0.2cm}
\begin{tabular}{lll}
$f(\phi)$ & $\left(p_\phi/\rho_\phi\right)_{today}$ & $\left(w_\phi\right)_{today}$ \\
\hline
$\phi^{-1/2}$ & -0.9967  &   -0.9973    \\
$\phi^{1/2}$  &  -0.9986 &    -0.9993     \\
$\phi^{-3/2}$ & -0.9947  &   -0.9953     \\
$\phi^{3/2}$  &  -1.0006 &    -1.0013     \\
$\phi^{-1}$   &   -0.9957&     -0.9963 \\
$\phi^{2}$    &    -1.0016&     -1.0023\\
$\phi^{-2}$   &   -0.9937&     -0.9943     \\
$\phi^{4}$    &    -1.0056&     -1.0063\\
$\phi^{-4}$   &   -0.9897&     -0.9903     \\
$\phi^{6}$    &    -1.0096&     -1.0103    \\
$\phi^{-6}$   &   -0.9857&     -0.9863     \\
$e^{\phi}$    &    -0.9997&     -1.0003\\
$e^{-\phi}$   &   -0.9957&     -0.9963\\
$\cos(\phi)+1$&-0.9966   & -0.9972
\\\hline \hline
\end{tabular}\end{center} \label{T1}
\end{table}

\subsection{The meaning of $w_\phi$}

Is $w_\phi$ representing an equation of state in the usual sense?
In other words, is the equation $\dot{\tilde \rho}_{\phi}=-3H(1+
w_{\phi})\tilde \rho_{\phi}\ ,$  satisfied? To answer
this question we shall rewrite the field equations for a general
non-minimally coupled theory as \cite{per2} \be G_{\mu\nu}=
\tilde T_{\mu\nu}, \label{CC} \ee where we have {\it defined} a
new stress-energy momentum tensor, on purpose, to make the
previous equation valid:
\begin{equation}
\tilde T_{\mu\nu}=T_{\mu \nu}^{fluid}+ \tilde T_{\mu\nu}[\phi]\ .
\end{equation}
As a consequence of the contracted Bianchi identities $\tilde
T_{\mu\nu}$ is conserved. And since there are no explicit coupling
between matter and fields, their corresponding energy-momentum
tensors are also separately conserved:
\begin{equation}
\nabla^{\mu}T_{\mu\nu}^{fluid}=\nabla^{\mu} \tilde T_{\mu\nu}[\phi
]=0\ .
\end{equation}
In this framework, reporting no more than a rewriting of the field
equations i.e. introducing no new physics at all, the explicit
expression for $\tilde T_{\mu \nu}[\phi]$ is given by
\begin{eqnarray}
\tilde T_{\mu \nu}[\phi] = \omega\left[\nabla_{\mu}\phi
\nabla_{\nu} \phi - {1 \over 2 } g_{\mu
\nu}\nabla^{\lambda}\phi\nabla_{\lambda}\phi\right]- \nonumber \\
Vg_{\mu \nu}+ \nabla_{\mu}\nabla_{\nu}F-g_{\mu \nu} \Delta F +
\left( 1-F\right) G_{\mu \nu} \ .
\end{eqnarray}
In order to obtain Eq. (\ref{CC}) we have just added and
subtracted $G_{\mu\nu}$ on the left hand side of the generalized
Einstein equations, i.e. $$ G_{\mu\nu} F= T_{\mu\nu}^{fluid} +
T_{\mu\nu}^{field},$$ with the usually defined
\begin{eqnarray}T_{\mu\nu}^{field}= \omega\left[\nabla_{\mu}\phi
\nabla_{\nu} \phi - {1 \over 2 } g_{\mu
\nu}\nabla^{\lambda}\phi\nabla_{\lambda}\phi\right]- \nonumber \\
Vg_{\mu \nu}+ \nabla_{\mu}\nabla_{\nu}F-g_{\mu \nu} \Delta
F.\end{eqnarray} This immediately fixes the expression of $\tilde
T_{\mu \nu}[\phi]$ above. The new effective energy density and
pressure that this defined stress-energy density produces are
\begin{equation}
\label{r} \tilde \rho_{\phi}= \omega {{\dot{\phi}}^2 \over 2
}+V(\phi)- 3 H\dot{F} + 3{H^2} \left( 1-F \right)\ ,
\end{equation}
\begin{equation}
\label{p} \tilde p_{\phi}= \omega {\dot{\phi}^2 \over 2
}-V(\phi)+2{H}\dot{F}+\ddot{F}- (3 H^2 + 2 \dot H ) \left(1
-F\right)\ ,
\end{equation}

In the case of Brans-Dicke theory, we recall,
$\omega=\omega_{BD}/\phi$ and $F=\phi$. Note then that $\tilde
\rho_\phi = \rho_\phi + 3H^2 (1-F)$, where $\rho_\phi$ was given
in Eq.~(\ref{f6}). The defined equation of state, $w_\phi$, thus,
is exactly that given by $\tilde p_\phi/\tilde \rho_\phi$, since
it was defined using the same Friedmann equation
$H^2=1/3(\rho_{fluid}+\tilde \rho_\phi)$. Indeed,
\begin{eqnarray}
\tilde \rho_\phi &=& - \rho_{fluid} + \frac 1F (\rho_{fluid} +
\rho_\phi) \nonumber \\
&=& - \rho_{fluid} + 3H^2 \nonumber \\
&=& - (3H^2F-\rho_\phi) + 3H^2 \nonumber \\
&=& \rho_\phi + 3H^2 (1-F)
\end{eqnarray}
At the same time, we can see that \be \tilde p_\phi=p_\phi - (3
H^2 + 2 \dot H ) \left(1 -F\right).\ee  {\it We conclude that the
definition for $w_\phi$ represents a real equation of state, since
it is supported by a conservation law, and that it is this the one
that should be taken into account to compare with the predictions
of GR, since it is directly related to the observable, $H$.} We
can also see, from Table I, that the difference between $w_\phi$
and the ratio $p_\phi/\rho_\phi$ is very small. The reasons that
leads to this are explicitly discussed for NMC theories in Section
III, a similar argument applies here as well.

\subsection{CMB-related observables}

The evolution of the comoving distance from the surface of last
scattering (defined as $z=1000$, equivalently $\ln a=-6.90$) can
be computed, for different theories, as: \be \int d\tau=
\int_{0.001}^a\;\; \frac{da}{a^2H(a)}.\ee Only in the case of
extremely low coupling factors (e.g. $\omega$ of Brans-Dicke
theory), discarded by current constraints, we see a noticeable
difference with the result of General Relativity plus cosmological
constant. To give a quantitative idea we can quote the ratio $$
\frac{\tau_{BD}-\tau_{GR}}{\tau_{GR}},  $$ calculated today
($a=1$),  which, for $\omega=25$ results equal to -0.014, whereas
for $\omega=500$ is $-5.7 \; 10^{-4}$, and quickly tends to zero
for bigger values of $\omega$.

The angular scales at which acoustic oscillations occur are
directly proportional to the size of the CMB sound horizon at
decoupling, that in comoving coordinates is roughly
$\tau_{dec}/\sqrt{3}$, and inversely proportional to the comoving
distance covered by CMB photons from last scattering until
observation, that is $\tau_{0}-\tau_{dec}$ \cite{Hu1}. The
multipoles scale as the inverse of the corresponding angular
scale, and so
\begin{equation}
 \ell_{peak}\propto
{\tau_{0}-\tau_{dec}\over\tau_{dec}}\ .
\end{equation}
As in the non-minimally coupled models studied in Ref.
\cite{bacci2}, $\tau$ changes because of a different dependence of
the Hubble length $H^{-1}$ in the past. However, we have already
noted that this change is almost imperceptible when compared with
usual General Relativity plus a cosmological constant, unless of
course (violating current constraints) the coupling parameter
$\omega$ is low enough.

The Integrated Sachs-Wolfe effect makes the CMB coefficients on
large scales, small $\ell$'s, change with the variation of the
gravitational potential along the CMB photon trajectories
\cite{Hu}. This is undoubtedly changed because of a variation in
the gravitational constant since the time of decoupling. However,
we expect this change to be also very small, since the
$G$-variation we have found, for values of $\omega=500$ and
bigger, are typically less than 2\% since the time of decoupling.

The scale entering the Hubble horizon at the matter-radiation
equivalence is also important, since it will define the matter
power turnover \cite{Coble}. The shift in the power spectrum
turnover is given by \cite{bacci2}
\begin{equation}
{\delta k_{turn}\over k_{turn}}=- \left({\delta H^{-1}\over
H^{-1}}\right)_{eq}\ ,
\end{equation}
and again, this reports a very small difference for all currently
possible $\omega$-values. Only for $\omega=25$ this difference is
about 12\% (where a case of a power law potential with exponent
equal to -2 taking as an example). For $\omega=500$ and bigger,
the differences are less than 1\% (to give a precise example it
reports a 0.7\% difference in the same power law case as commented
before and $\omega=500$). Contrary to what Baccigalupi et al. have
done in the past \cite{perrota} , we are not comparing two
different theories (Brans-Dicke and General Relativity) with the
same potential, but rather, and motivated by the findings of this
section, Brans-Dicke theory with any given potential and General
Relativity plus $\Lambda$. It is in this case that the
possibilities of actually distinguishing both situations diminish.


\section{Non-minimally coupled Theories}

The details of the cosmological evolution using the general action
(\ref{action}) above, with $\omega=1$ and \be F(\phi)= 1 + \xi
\phi^2 \ee were explored in Refs. \cite{perrota,per2,bacci2},
among many others, and we refer the reader to these works and
references therein for additional relevant discussions. Our
numerical code is in agreement with the results therein reported,
and it is a direct extension of the numerical implementation
reported in Section IIIB. \\

\begin{figure}
\includegraphics[width=9cm,height=12cm]{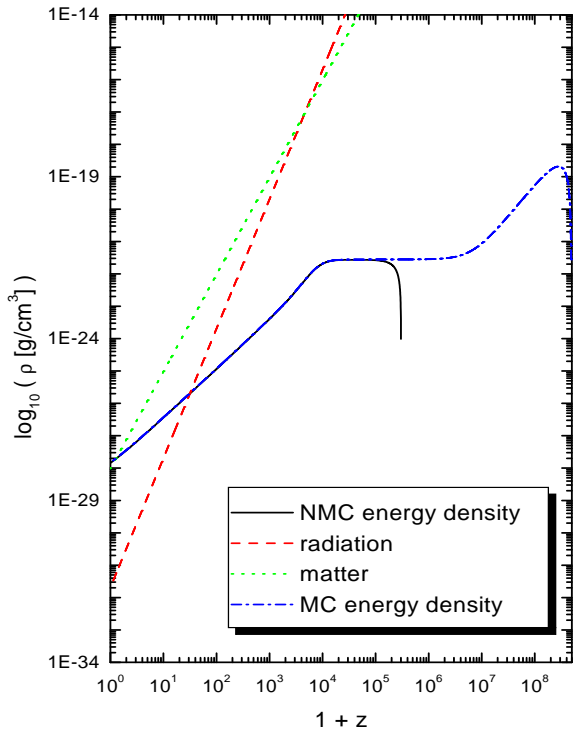}
\caption{Matter, radiation and quintessence energy densities, both
non-minimally (NMC) and minimally coupled (MC). The incomplete
curve corresponds to the non-minimally coupled quintessence, and
the cutoff is produced by a change in the sign of the effective
density, see text for details.  } \label{om-nmc1}
\end{figure}

In this Section, following the previous discussion, we would like
to focus on the possible definitions of {\sl equations of state}
and their impact onto observable quantities. Just as an example,
we show in Figure \ref{om-nmc1} the case of a tracking potential
of the form $V=V_0 \phi^{-2}$, $H_0=70$km/s/Mpc, in a flat
universe with $\Omega_{m,0}=0.4$. The equivalent Brans-Dicke
parameter (obtained by redefining fields in Eq. (\ref{action}) in
order for it to look like a Brans-Dicke theory) is
$\omega_{JBD}=\omega F/F_{,\phi}^{2}$, and its value is given by
defining $\xi$. The value of $\xi$ used in the model of Figure
\ref{om-nmc1} and successive ones is 5.8 $10^{-3}$, what implies
an equivalent Brans-Dicke parameter $\omega_{BD}=3071$, well in
agreement with current constraints.

Starting from the general field equations, we can immediately
define, as done for the Brans-Dicke theory, an effective energy
density and pressure for the scalar field. The former, for
instance, appears writing the 00-component of the field equation
as $ H^2= 1/{3F} \left( \rho_{fluid} + \rho_\phi \right). $ The
explicit expression then being, \be \rho_\phi={\omega \over 2}
\dot{\phi}^2
 +V - 3 { H} \dot{F} \ee for the energy
density, and \be p_\phi= \frac \omega2 \dot \phi^2 -V+2 H \dot F +
\ddot F \ee for the pressure. These two expressions do not, as we
have shown before, pertain to a conserved energy-momentum tensor.
They do, however, define an effective equation of state, this
being just $p_\phi/\rho_\phi$. This relationship is subject to
same caveats mentioned above for the case of Brans-Dicke: it is
neither positive nor negative defined, since the effective energy
density itself shifts its sign during the evolution. The energy
density quoted above is what is depicted (whenever possible) in
Figure \ref{om-nmc1}. As it can be seen, it tracks closely the
usually defined minimally coupled (MC) energy density at low
redshifts, this being an effect of the necessarily small coupling
$\xi$ that is adopted to fulfill observational constraints.

Again, in order to work with a conserved energy-momentum tensor,
we can rewrite the field equations and obtain a real equation of
state, $w_\phi=\tilde p_\phi/\tilde \rho_\phi$, where $\tilde
\rho_\phi$ and $\tilde p_\phi$ were given in Eqs. (\ref{r}) and
(\ref{p}), respectively. As we already mentioned in the case of
Brans-Dicke, this equation of state is exactly what results in
writing the field  equation as $H^2=1/3 ( \rho_{fluid} +
\varrho_\phi)$, defining implicitly $\varrho_\phi=1/F
(\rho_{fluid}+\rho_\phi)-\rho_{fluid}.$ Finally, just for
comparison, one can as well consider the equation of state for the
case of a minimally coupled scalar field, $p/\rho$, with
\begin{equation}
 \rho={1\over 2}\dot{\phi}^{2}+V(\phi )\ ,
\end{equation}
\begin{equation}
p={1 \over 2}\dot{\phi}^{2}-V(\phi )\ .
\end{equation}


\begin{figure*}
\includegraphics[width=8cm,height=12cm]{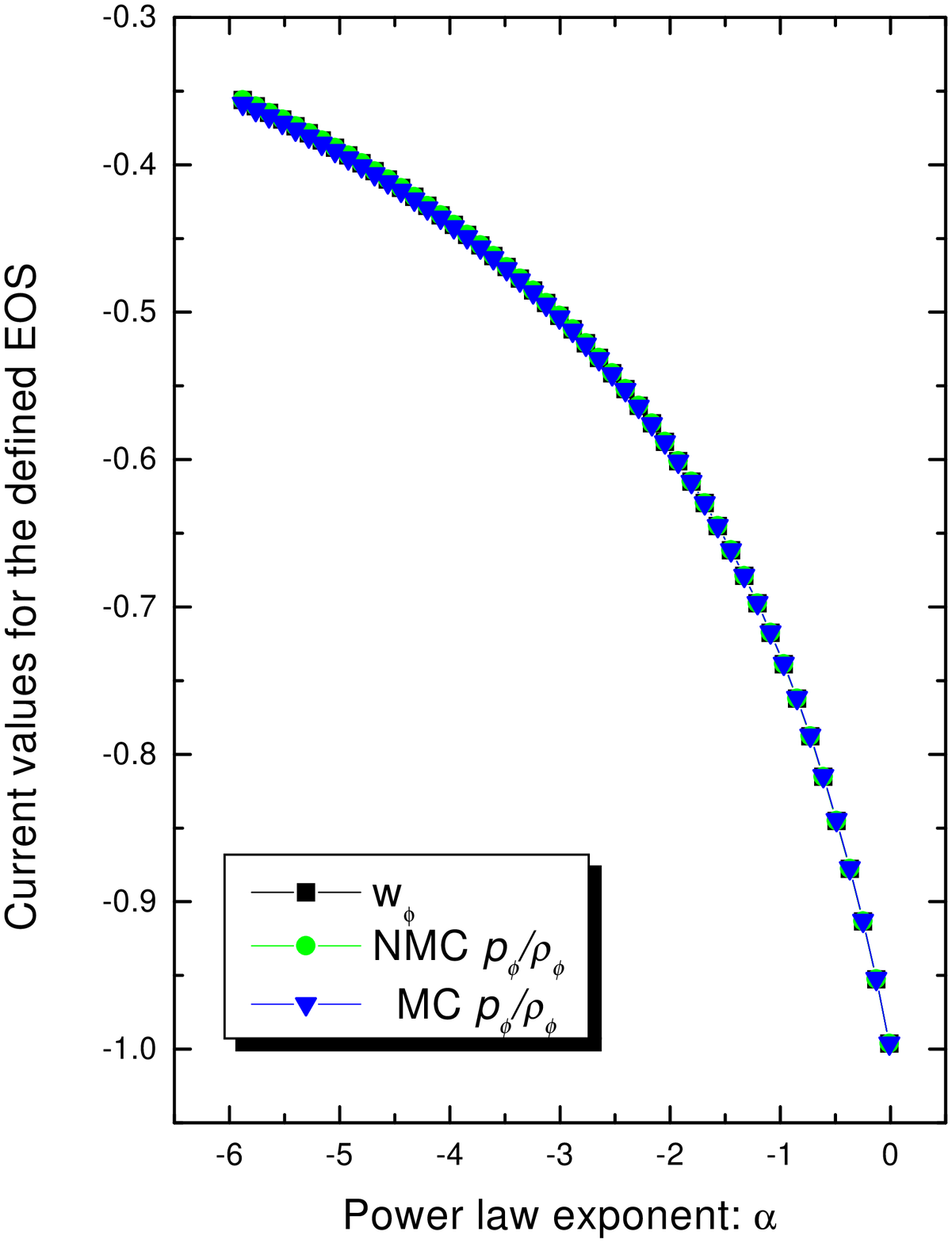}
\includegraphics[width=8cm,height=12cm]{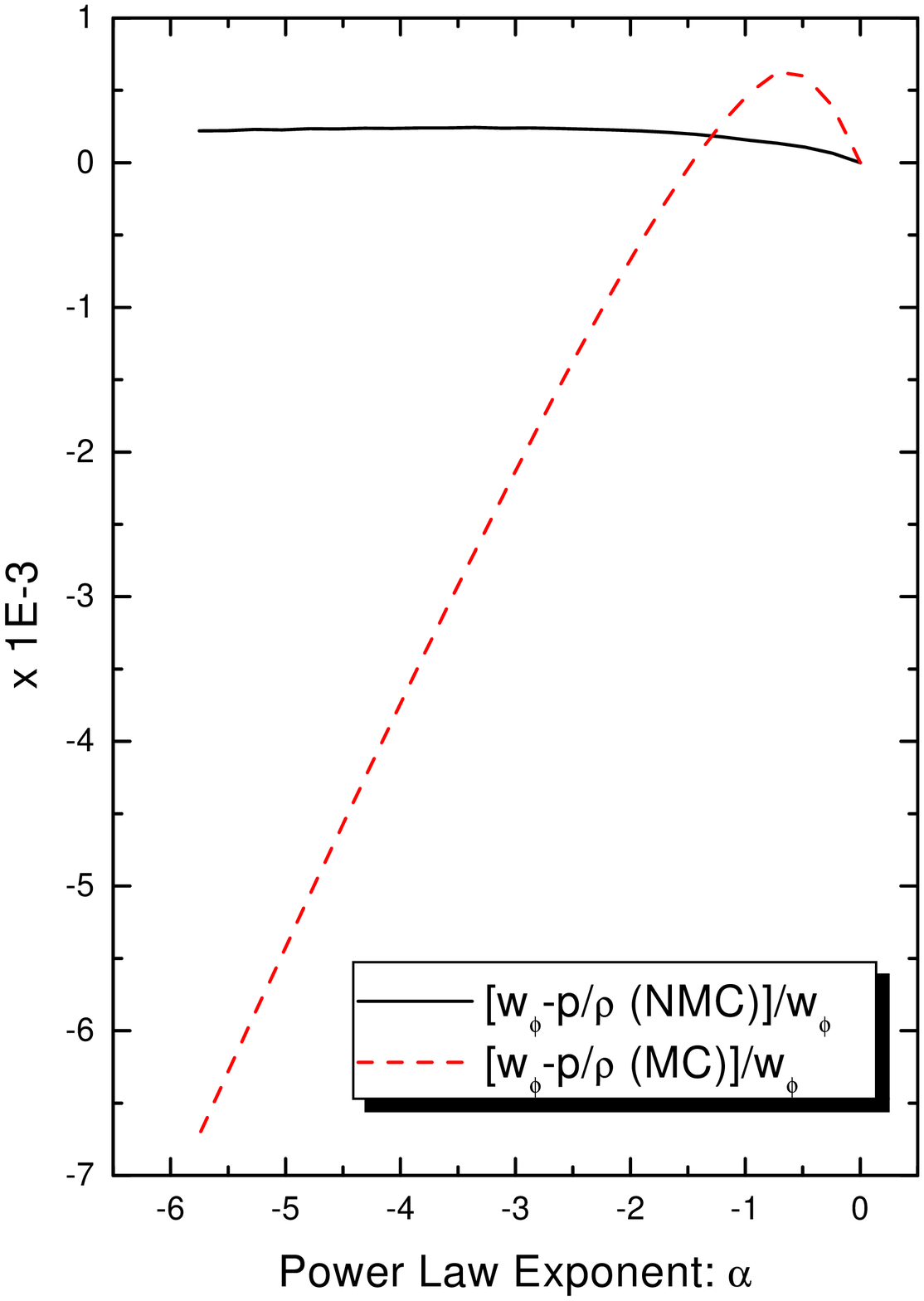}
\caption{Left: Current values of the defined 'equations of state',
see text, as a function of the power law exponent of the tracking
potential. Right: Difference between the equation of state
directly related with the Hubble parameter, $w_\phi$ and the
non-minimally coupled (NMC) and minimally coupled
$p_\phi/\rho_\phi$. } \label{eos-today}
\end{figure*}

In Figure \ref{eos-today}  we show the results of these different
definitions of equation of state for the current time, as a
function of the exponent of the power law potential
$V=V_0\,\phi^\alpha$, for a flat universe model given by
$\Omega_{m,0}=0.4$, the same model used in Figure \ref{om-nmc1}.
We see that they do not present noticeable differences. Very low
values of the power law exponent (shallow potentials) are needed
to produce equations of state near that generated by a
cosmological constant.  To be quantitative, Figure \ref{eos-today}
also shows, in the right panel, the differences between the
equation of state directly related with the Hubble parameter,
$w_\phi$ and the non-minimally coupled (NMC) and minimally coupled
$p_\phi/\rho_\phi$. Clearly, the differences are minor. One can
actually understand why these differences are so small. Note that
the energy density and pressure in a non-minimally coupled field
theory can be written as the minimally coupled ones plus
additional terms. In the case of $\tilde \rho_\phi$, these terms
are equal to $-3H\dot F + 3H^2(1-F)$, whereas for $\rho_\phi$ only
the first term above enters. Both these terms are, however,
proportional to $\xi$, being themselves $$ -3H\dot F + 3H^2(1-F)
\sim -6 \xi H\dot \phi - 3\xi H^2  \phi^2.$$ But since from the
evolution of the field, ${\cal O}(H^{-1} \dot \phi)$=1 today, and
at the current era, ${\cal O}(H^2) \sim V$, the energy density can
be written as $\tilde \rho_\phi \sim \rho_{mc} -3 \xi (\frac{\dot
\phi^2}{2}+V\phi^2) $, where $\rho_{mc}$ is the minimally coupled
energy density. Clearly, at the current cosmological era and
because of the constraints on $\xi$, the second terms are
sub-dominant in comparison to the first ones. A similar analysis
can be established for the pressure, where again all extra terms
are proportional to $\xi$, and then for the equations of state.
Today the influence of the last terms in Eqs. (\ref{r}) and
(\ref{p}), the `gravitational dragging' terms, as dubbed in Ref.
\cite{per2}, is negligible in comparison to the minimally coupled
contribution. It is only in the past, when the matter density
dominates the evolution of the universe, that these terms become
important.

\begin{figure}
\includegraphics[width=9cm,height=12cm]{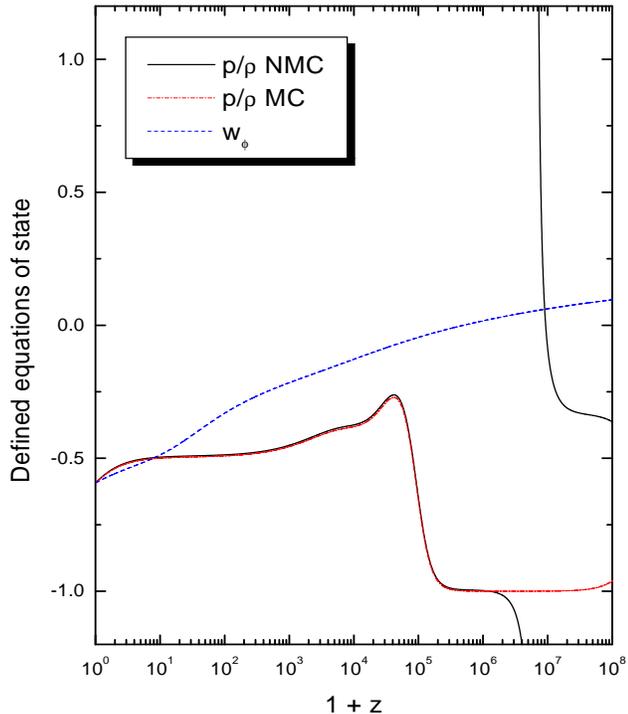}
 \caption{Evolution in redshift of the equations of state
for a model with $V\propto \phi^{-2}$. } \label{eos-today2}
\end{figure}

The previous analysis is not valid in the past history of the
universe. Figure \ref{eos-today2} shows the evolution of the
different equations of state with redshift. We can see that the
scaling solution of the tracking potentials (equation of state is
approximately equal to $-2/(2+\alpha)$ between $z\sim$1--1000,
when $\rho_{matter}>\rho_\phi$, see for e.g. Refs.
\cite{perrota,per2}), appears for all defined equations of state
but $w_\phi$. We also note that because of the sign
non-definiteness of the defined non-minimally coupled effective
`energy density', the corresponding equation of state shows sudden
changes at high redshift, in the position where the density cross
zero. Finally, we find that for tracking non-minimally coupled
quintessence, {\it there is no super-quintessence potential in any
of the defined equations of state, not even by an small amount},
all of them being greater than -1 at present. This latter result
is actually valid for all potentials analyzed (similar cases than
those presented in Table I).

\subsection{Comments on perturbations and on the possible degeneracy with
kinetically driven quintessence}

Perturbations analysis of the field equations, as done by Perrotta
and Baccigalupi \cite{per2}, have shown that rich phenomena are
uncovered when working with the field equations written as in
(\ref{CC}). The most important of them are those generated by
gravitational dragging, exclusive of non-minimally coupled
theories. This phenomenon is, basically, the early dominance of
the last term in Eqs. (\ref{r}) and (\ref{p}), which is, in turn,
produced due to the fact that they are proportional to the square
of the Hubble parameter, and then to the total energy density. The
latter scales as $a^{-3}$ and $a^{-4}$ at early times, and then
makes the gravitational generated term to dominate the dynamics of
the field.

The advantage of writing the formalism as in Eq. (\ref{CC}) is the
fact that usual perturbation analysis --as applied for any fluid
component of the universe-- is also valid for the field. In that
sense, concepts as the equation of state, or the adiabatic sound
speed \be c^2_\phi=\frac{\dot{\tilde p}_{\phi}}{\dot{\tilde
\rho}_{\phi}} = w_\phi  - \frac{1}{3{ H}} \frac {\dot
w_\phi}{1+w_\phi}, \ee (where we have made use of the field
conservation equation) can be well defined. If the field is slowly
varying in time, $c^2_\phi \sim w_\phi$. Hu \cite{Hu} showed, for
negative equations of state, that adiabatic fluctuations are
unable to give support against gravitational collapse. Density
perturbations would become non-linear after entering the horizon,
unless the entropic term $w_\phi \Gamma_\phi>0$ \cite{Hu}.

The effective sound speed, $c_{{{\rm eff}}\,,\phi}^2$, is then
defined in the rest frame of the scalar field, where $\delta
T^0_{j\,\phi}=0$ \cite{Hu}. The gauge invariant entropic term is
written as $ w_{\phi} \Gamma_{\phi}=(c_{{\rm eff}}^2-c_{\phi}^2)
\delta_{\phi}^{(rest)}\ , $ where $\delta_{\phi}^{(rest)}$ is the
density contrast in the dark energy rest frame \cite{per2} $
\delta_{\phi}^{(rest)}= \delta_{\phi}+3{{\cal{H}} \over
k}\left(1+w_{\phi}\right)\left(v_{\phi}-B\right) $ (we refer the
reader to Ref. \cite{per2} for more careful explanations). For
normal quintessence, the effective sound speed is $c_{{{\rm
eff}}\,,\phi}^2=1$, giving a relativistic behavior to the
corresponding density fluctuations. However, Perrotta and
Baccigalupi \cite{per2} have found that the situation can be much
different for non-minimally coupled gravity. In that case,
\begin{equation}
c_{{{\rm eff}}\,,\phi}^2
\sim {\delta \tilde
p_{\phi} \over \delta \tilde \rho_{\phi}}
\end{equation}
for values of $k\gg { H}$. But because of gravitational dragging,
$\delta \tilde \rho_{\phi}$ can be quite difference from the usual
case,  and this ratio may be much lower than unity whenever the
energy density perturbations of the scalar field are enhanced by
perturbations in the matter field. At the level of perturbations,
then, quite distinctive effects appears in non-minimally coupled
quintessence as compared with the usual case and make these
theories possibly distinguishable.

Very recently, yet another scenario for an alternative model of
quintessence was introduced \cite{picon}. In it, known as
$k$-essence, the Lagrangian density includes a non-canonic kinetic
term: \be S = \int d^4x\sqrt{-g}\left({1\over 16\pi
G}R+p(\p,\nabla
  \p)\right)+S_{m,r},
\ee where $S_{m,r}$ denotes the action for matter and/or
radiation. Examples in which the Lagrangian depends only on the
scalar field $\phi$ and its derivative squared $ X =-{1\over
2}\nabla^{\mu}\p\nabla_{\mu}\p. $ have been constructed
\cite{picon}. The field equations for this models are
\begin{eqnarray}  R_{\mu\nu}-{1\over 2}g_{\mu\nu}R  =
\;\;\;\;\;\;\;\;\;\;\;\; \;\;\;\;\;\;\;\;\;\;\;\;\;
\;\;\;\;\;\;\;\;\;\;\;\;&&  \nonumber
\\
8\pi G \left(\frac{\partial p(\p,X)}{\partial X}\nabla_{\mu}\p
\nabla_{\nu}\p+p(\p,X)g_{\mu\nu}+T^{m,r}_{\mu\nu}\right)
\end{eqnarray} where $T^{m,r}_{\mu\nu}$ is the
energy-momentum tensor for usual matter fields. $p(\p,X)$
corresponds to the pressure of the scalar field, whereas the
energy density is given by $\rho_{\p}=2X\partial p/\partial X-p$
\cite{picon2}. It can be seen that for this models, the speed of
sound is given by \cite{erik} \be c_{{\rm eff}}^{2}={p_{Q,X}\over
\rho_{Q,X}}={p_{,X}\over p_{,X}+2Xp_{,XX}}. \ee Apparently, then,
and since $p$ is completely generic, it could exist a non-canonic
kinetic term within $k$-essence giving rise to the same results of
non-minimally coupled gravity. Viscosity (a parameter relating
velocity and metric shear with anisotropic stress \cite{Hu}) can
however provide the way to break the degeneracy, since it results
non-zero for a non-minimally coupled field (contrary as well to
what results in usual quintessence) \cite{per2}.

\section{Conclusions}

In the case of Brans-Dicke theory, and in the cases of
$\omega_{BD}$ allowed by current constraints, we have numerically
proven that the homogeneous field equations of extended
quintessence yield to no observable effect that can distinguish
the theory from the predictions of General Relativity plus a
cosmological constant. It is with this model that the comparison
should be made, since for the large values of the coupling
parameters required by current experiments, all potentials are
closely similar to a constant function, and the theory itself to
General Relativity. Although we have not made a detailed
perturbation analysis using the full numerical CMBFAST code, we
can safely predict that the same situation will happen there, as a
result of the analysis made for the CMB-related observables in
Section IIC. We discussed the observationally-related definition
of equation of state $w_\phi$, and not to the usually studied
ratio between the effective pressure and density directly obtained
from the field equations, to which we assign no particular
physical meaning. The phenomenon of super-quintessence, i.e. a
super-accelerated expansion of the universe, although possible for
a non-minimally coupled Brans-Dicke scalar field, and impossible
in any minimally coupled field situation, it is of such an small
amount that is far beyond the expectations of any realistic
experiment. From a practical point of view, then,  it will always
exist a scalar field potential supporting a minimally coupled
field that produces experimentally indistinguishable results from
those obtained within the extended quintessence framework of
Brans-Dicke theory.

\mbox{} For the more general extended non-minimally coupled models
studied, the possibility of having super-quintessence actually
disappears: all tracking potentials explored produce effective
equations of state greater than -1. We have shown that for low
values of the exponent in the tracking potentials supporting the
non-minimally coupled field (i.e. equations of state are close to
-1), the difference among all defined equations of state is
negligible. It is, however, in the perturbation regime where
differences with the usual quintessence case can be noticed. As
Perrotta and Baccigalupi have found \cite{per2}, a new
gravitational dragging effect appears here, giving rise to the
possibility of clumps of scalar field matter. In this case,
however, it is with $k$-essence models that a degeneracy could
appear, particularly in those cases in which $p_{,X}+2Xp_{,XX} \gg
1$, yielding the speed of sound to values close to zero.

Finally, we remark that expanding solutions where acceleration is
transient have been recently considered given the consistency
problem between string theory and spaces with future horizons
\cite{hellerman}. Since scalar-tensor theories of gravity likely
originate in string theory, it would be interesting to make a
similar analysis to that presented in the previous reference for
the case of non-minimally coupled theories.

\acknowledgments It is a pleasure to warmly acknowledge Prof. Uros
Seljak.  His contribution and permanent advice were invaluable.
Very useful discussions with Dr. F. Perrotta, as well as
interesting comments by Dr. A. Mazumdar, are also thankfully
acknowledged. Dr. Perrotta is further acknowledged for his
critical reading of the manuscript. An important improvement of
the paper has been possible after useful remarks from an anonymous
Referee.

\subsection*{Appendix}

In the literature, one may find two alternative introductions of
general non-minimally coupled theories. Firstly, the one that we
follow in Section II (see for instance \cite{PRL}), and secondly,
the one that is derived from the action
\begin{eqnarray}
 S=\int d^4 x \sqrt{-g} \left[ {1 \over 2\kappa}
f(\phi, R) - {\omega (\phi )\over 2} \nabla^{\mu} \phi
\nabla_{\mu}\phi \right. \nonumber \\ \left. -V( \phi) +
L_{fluid}\right]\ ,
\end{eqnarray}
where $\kappa$ is a constant, not necessarily taken as 1, and
plays the role of the ``bare" gravitational constant (see for
instance \cite{farese}, by the same authors). The function $f$ is
then assumed to be of the form
\be(1/\kappa)f(\phi,R)=F(\phi)\,R.\ee The function $F$, in the
case we are interested in, is written as \be F(\phi)=\frac 1\kappa
+ \xi \phi^2 = 1 + \xi (\phi^2-\phi_0^2) .\ee Then, a value of
$\kappa=1$, as we have taken in the theoretical development of the
previous sections just reduces to take $\phi_0$, the current value
of the field, equal to 0. This, however, is not what may result,
in general, numerically convenient, since it would imply to
precisely fix the evolution of the field to reach $\phi=0$ today.
Instead, as we are not actually interested in any value of
$\kappa$ per se, we do not make $\kappa=1$ in our numerical code,
and instead follow the treatment given by Perrotta and Baccigalupi
(\cite{per2}). In that case, they choose to rewrite the field
equations as \be \label{la} G_{\mu\nu}=\kappa T_{\mu\nu},\ee with
the corresponding field energy density and pressure given by
\begin{equation}
\label{r2} \tilde \rho_{\phi}= \omega {{\dot{\phi}}^2 \over 2
}+V(\phi)- 3 H\dot{F} + 3{H^2} \left( \frac 1\kappa -F \right)\ ,
\end{equation}
\begin{equation}
\label{p2} \tilde p_{\phi}= \omega {\dot{\phi}^2 \over 2
}-V(\phi)+2{H}\dot{F}+\ddot{F}- (3 H^2 + 2 \dot H ) \left( \frac
1\kappa -F\right)\ .
\end{equation}
When comparing Eq. (\ref{la}) with the usual Einstein equations,
one has to take into account that the value of $\kappa$ is not 1
(although certainly it is truly close to unity, because of the
constraint imposed on $\xi$). Then, if we decide to write the
Friedman equation like $H^2= 1/3 (\rho_{fluid} + \varrho_\phi)$,
to directly compare with GR (and the same matter density) the
corresponding relationship between $\varrho_\phi$ and $\tilde
\rho_\phi$ give in Eq. (\ref{r2}) is \be \varrho_\phi = (\kappa -
1) \rho_{fluid} + \kappa \tilde \rho_\phi . \ee In this scheme,
$\tilde p_\phi/\tilde \rho_\phi$ (with quantities defined as in
Eqs. (\ref{r2}) and (\ref{p2})) will differ from the equation \be
\label{extra2} w_\phi=-1-\frac 13 \ln \left(\frac{
\varrho_\phi}{\varrho_{\phi\,0}}\right)\frac 1{\ln (a)},\ee
because $\varrho_\phi \neq \tilde \rho_\phi$, but it is the latter
Eq. (\ref{extra2}) what should be used to compare with the results
of General Relativity with a fixed current matter density.



\end{document}